\documentclass[doublecol]{epl2} 

\usepackage{amsmath,amssymb}
\usepackage{epstopdf}
\usepackage{bm}
\title{Anomalous lateral diffusion in a viscous membrane surrounded by 
viscoelastic media}
\shorttitle{} 

\author{
S. Komura\inst{1}\thanks{E-mail: \email{komura@tmu.ac.jp}}, 
S. Ramachandran\inst{2}
\and 
K. Seki\inst{3}
}

\shortauthor{S. Komura, S. Ramachandran, K. Seki}

\institute{                    
\inst{1}
Department of Chemistry, 
Graduate School of Science and Engineering, 
Tokyo Metropolitan University, 
Tokyo 192-0397, Japan \\
\inst{2} 
Physique des Polym\`eres, Universit\'e Libre de Bruxelles,
Campus Plaine, CP 223, B-1050 Brussels, Belgium \\
\inst{3}
National Institute of Advanced Industrial Science and Technology (AIST),  
Ibaraki 305-8565, Japan
}


\abstract{
We investigate the lateral dynamics in a purely viscous lipid membrane 
surrounded by viscoelastic media such as polymeric solutions. 
We first obtain the generalized frequency-dependent mobility tensor 
and focus on the case when the solvent is sandwiched by hard walls.
Due to the viscoelasticity of the solvent, the mean square displacement 
of a disk embedded in the membrane exhibits an anomalous diffusion.
An useful relation which connects the mean square displacement and the 
solvent modulus is provided.
We also calculate the cross-correlation of the particle displacements which 
can be applied for two-particle tracking experiments.
}

\begin{document}

\maketitle

\section{Introduction}

Biomembranes are thin two-dimensional (2D) fluids which separate 
inner and outer environments of organelles in cells.
The fluidity of biomembranes is guaranteed mainly due to the lipid 
molecules which are in the liquid crystalline state at physiological 
temperatures. 
Proteins and other molecules embedded in biomembranes undergo lateral 
diffusion which plays important roles for biological 
functions~\cite{AlbertsBook}.    
It should be noted, however, that biomembranes are not isolated 2D systems, 
but are coupled to the surrounding polar solvent such as water.   
Indeed the presence of water is essential for amphiphilic lipid molecules 
to spontaneously form bilayers by self-assembly.

In the last few decades, it has been recognized that the outer solvent 
has a significant effect on the membrane dynamics which takes place in 2D. 
Saffman and Delbr\"uck considered the Brownian motion of a 2D disk confined 
in a fluid membrane~\cite{saffman-75,saffman-76}.
In their hydrodynamic model, the transfer of membrane momentum to the 
outer three-dimensional (3D) fluid was taken into account through the 
boundary conditions at the membrane surfaces.  
The translational diffusion coefficient in the weak coupling (small disk) 
limit was shown to exhibit a logarithmic dependence on the disk size.
This dependence has been repeatedly tested for various 
lipid molecules and proteins~\cite{Gambin-06}.
In the strong coupling (large disk) limit~\cite{hughes-81}, on the other 
hand, the diffusion coefficient is inversely proportional to the disk 
size showing the analogy to the 3D  Stokes-Einstein 
relation~\cite{LandauBook}.
Such a 3D-like behavior in 2D membrane is caused by the back flow effect 
mediated by the bulk solvent.

In this Letter, we discuss the dynamics and responses of membranes when 
their surrounding solvent is viscoelastic rather than purely viscous.
This is a common situation in all eukaryotic cells whose cytoplasm
is a soup of proteins and organelles, including a thick sub-membrane
layer of actin-meshwork forming a part of the cell 
cytoskeleton~\cite{AlbertsBook}.
The extra-cellular fluid can also be viscoelastic because it is filled 
with extracellular matrix or hyaluronic acid gel.
In addition to the surrounding solvents, lipid membranes themselves can be 
viscoelastic~\cite{HBP10}. 
Although it turned out that membranes are purely viscous in the latest 
report~\cite{HBP10}, their experimental technique using particle tracking 
microrheology provides us with a new clue to investigate the dynamical 
responses of lipid bilayers coupled to the surrounding environments 
under controlled conditions.
Recently, viscoelasticity of phospholipid Langmuir monolayers in a 
liquid-condensed phase was measured using active microrheology~\cite{Choi11}. 
Being motivated by these works, we discuss the mean square displacement 
(MSD) of a circular disk embedded in a 2D sheet by taking into account the 
viscoelasticity of the surrounding media.
This quantity is experimentally measurable in single-particle tracking 
microrheology~\cite{MW95,MGZWK}.
We further calculate the two-particle MSD which is
useful for two-point microrheology experiments~\cite{CVWGKYW}. 
For both cases, we show that the viscoelasticity of the surrounding media 
leads to an anomalous diffusion in the 2D viscous membrane.

Recently, Granek discussed the dynamics of an undulating bilayer membrane
surrounded by viscoelastic media~\cite{granek-11}.
He calculated the frequency-dependent transverse (out-of-plane) MSD of 
a membrane segment and the linear response to external forces.  
In our theory, we treat the membrane as an infinitely large flat sheet 
with a 2D viscosity $\eta$, and consider its lateral dynamics.
A similar problem was considered in refs.~\cite{CB11,HH11} in which 
the authors have taken into account the viscoelasticity of the membrane 
itself because they were originally motivated by the earlier experiment of 
ref.~\cite{HBP10}.
A more general theory for the dynamics of viscoelastic membranes was
given by Levine and MacKintosh~\cite{LM02}.
In these works, however, the surrounding solvent is assumed to be
purely viscous.
Similar to Granek's work, the main purpose of our work is to point 
out the importance of the viscoelasticity of the bulk solvent on the 
membrane lateral dynamics such as diffusion or linear viscoelastic 
response.    
To make this point clear enough, we intentionally treat the membrane
as a purely viscous 2D fluid.

\section{Hydrodynamic model}

\begin{figure}[t]
\begin{center}
\includegraphics[scale=0.4]{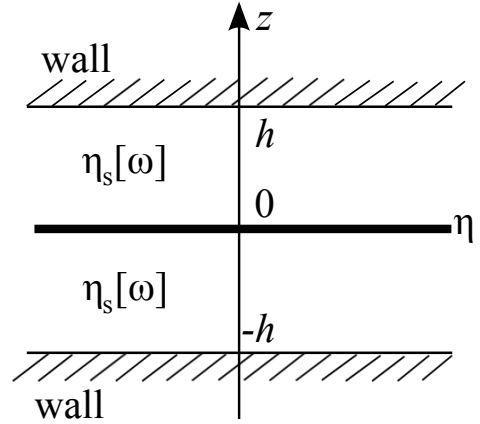}
\caption{Schematic picture showing a planar viscous membrane 
at $z=0$ with 2D constant viscosity $\eta$.
It is sandwiched by a viscoelastic solvent of 3D frequency-dependent 
viscosity $\eta_{\rm s}[\omega]$.
Two impenetrable walls are located at $z=\pm h$ bounding the solvent.}
\label{image}
\end{center}
\end{figure}

We first establish the governing hydrodynamic equations for our model.
As shown in fig.~\ref{image}, the fluid membrane, fixed in the $xy$-plane 
at $z=0$, is embedded in a bulk solvent that is further bounded by hard 
walls at $z=\pm h$. 
Let $\mathbf{v}(\mathbf{r},t)$ be the 2D velocity of the membrane fluid 
at position $\mathbf{r}=(x,y)$ and at time $t$.
We assume that the membrane is incompressible;
\begin{equation}
\nabla \cdot \mathbf{v}= 0.
\label{memincompress}
\end{equation}
We also work in the low-Reynolds number regime so that the inertial effects 
can be neglected. 
Then the Stokes equation for the fluid membrane is given by 
\begin{equation}
\rho \frac{\partial \mathbf{v}}{\partial t}=
\eta \nabla^2 \mathbf{v} -\nabla p
+\mathbf{f}_{\rm s}+\mathbf{F},
\label{memequation}
\end{equation}
where $\rho$ is the membrane 2D density, $\eta$ the membrane 2D 
constant viscosity, $p$ the in-plane pressure, 
$\mathbf{f}_{\rm s}$ the force due to the solvent described later, 
and $\mathbf{F}$ is any other force acting on the membrane.
Notice that $\nabla$ stands for the 2D differential operator.

Next we give the equations for the surrounding viscoelastic solvent.   
The upper ($z>0$) and the lower ($z<0$) regions of the solvent are denoted 
by ``$+$'' and ``$-$'', respectively.
The velocities and pressures in these regions are written as 
$\mathbf{v}^{\pm}(\mathbf{r},z,t)$ and $p^{\pm}(\mathbf{r},z,t)$, 
respectively.
Similar to the fluid membrane, the solvent is also assumed to be 
incompressible;
\begin{equation}
\tilde{\nabla} \cdot \mathbf{v}^{\pm}= 0.
\label{solincompress}
\end{equation}
The Stokes equation for the viscoelastic solvent is written as 
\begin{equation}
\rho_{\rm s} \frac{\partial \mathbf{v}^{\pm}}{\partial t}=
\tilde{\nabla} \cdot {\bm \sigma}^{\pm}-\tilde{\nabla} p^{\pm},
\label{solequation}
\end{equation}
where $\rho_{\rm s}$ is the solvent 3D density (assumed to be the same for
both solvents), and $\tilde{\nabla}$ indicates the 3D differential operator.
Within the linear viscoelasticity approximation, the stress tensor 
${\bm \sigma}^{\pm}$ in the above equation is given by the following 
constitutive relation~\cite{BirdBook}
\begin{equation}
{\bm \sigma}^{\pm}(t)= 2 \int_{-\infty}^{t} {\rm d}t' \,
\eta_{\rm s}(t - t') \mathbf{D}^{\pm}(t'), 
\label{eqn2:tau}
\end{equation}
where $\eta_{\rm s}(t)$ is the time-dependent solvent viscosity (assumed to 
be the same for both solvents).
The rate-of-strain tensor is given by
$\mathbf{D}^{\pm}=  
[ \tilde{\nabla} \mathbf{v}^{\pm} + 
(\tilde{\nabla} \mathbf{v}^{\pm})^{\rm T} ]/2$,
where the superscript ``${\rm T}$'' denotes the transpose.
The surrounding viscoelastic solvent exerts force $\mathbf{f}_{\rm s}$
on the membrane as considered in eq.~(\ref{memequation}).     
It is given by the projection of 
$({\bm \sigma}^{+} - {\bm \sigma}^{-})_{z=0} \cdot \hat{\mathbf{e}}_z$
on the $xy$-plane, where $\hat{\mathbf{e}}_z$ is the unit vector along 
the $z$-axis.
We employ the stick boundary conditions at any time, i.e., the equalities 
of the velocities, at $z=0$ and $z=\pm h$.
When the solvent is purely viscous, this model reduces to that
considered by the present authors~\cite{sanoop-poly-11}.

It is convenient to perform the 2D Fourier transform in space and 
the Fourier-Laplace (or one-sided Fourier) transform in time for any 
function $f(\mathbf{r},t)$ as defined by 
$f[\mathbf{k},\omega]= \int_{-\infty}^{\infty} {\rm d}^2 r \, 
\int_0^{\infty} {\rm d} t \,
f(\mathbf{r},t) \exp[ -i(\mathbf{k} \cdot \mathbf{r}+\omega t)]$,
where $\mathbf{k}=(k_x,k_y)$ is the 2D wavevector and $\omega$ the 
angular frequency.
Assuming that fluids are at rest at $t=0$, we calculate 
$\mathbf{f}_{\rm s}$ and obtain the membrane velocity as  
$\mathbf{v}[\mathbf{k},\omega]=
\mathbf{G}[\mathbf{k},\omega] \cdot 
\mathbf{F}[\mathbf{k},\omega]$,
where $\mathbf{G}[\mathbf{k},\omega]$ is the frequency-dependent mobility 
tensor.
Following the similar procedure described in refs.~\cite{CB11,HH11}, we 
obtain 
\begin{equation}
G_{\alpha\beta}[\mathbf{k},\omega]= 
\frac{\delta_{\alpha\beta}-k_\alpha k_\beta/k^2}
{\eta k^2 + 2 \eta_{\rm s}[\omega] k' \coth(k'h) + i \omega \rho},
\label{genoseen}
\end{equation}
where 
$k'= k \left( 1 + i \omega \rho_{\rm s}/\eta_{\rm s}[\omega] k^2 \right)^{1/2}$,
$k = \vert \mathbf{k} \vert$ and $\alpha, \beta = x, y$.
In the limit of $h \rightarrow \infty$, which we call as the 
``free membrane case'', the above mobility tensor reduces to  
\begin{equation}
G_{\alpha\beta}^{\rm free}[\mathbf{k},\omega]= 
\frac{\delta_{\alpha\beta}-k_\alpha k_\beta/k^2}
{\eta k^2 + 2 \eta_{\rm s}[\omega] 
k \sqrt{1+i \omega \rho_{\rm s}/\eta_{\rm s}[\omega] k^2} + 
i \omega \rho},
\label{freemem}
\end{equation}
which was given in ref.~\cite{CB11}.
The opposite limit of $h \rightarrow 0$ is called as the ``confined 
membrane case'' and the corresponding mobility tensor becomes  
\begin{equation}
G^{\rm con}_{\alpha\beta}[\mathbf{k},\omega]=
\frac{\delta_{\alpha\beta}-k_\alpha k_\beta/k^2}
{\eta( k^2 + \kappa^2[\omega])},
\label{conmem}
\end{equation}
where we have introduced $\kappa[\omega]$ whose square is defined as
\begin{equation}
\kappa^2[\omega] = \frac{2\eta_{\rm s}[\omega]}{\eta h} 
+ \frac{i \omega \rho}{\eta}.
\label{sqscreen}
\end{equation}

In the limit of $\omega \rightarrow 0$, eq.~(\ref{genoseen}) reduces 
to the static mobility tensor obtained in ref.~\cite{sanoop-poly-11}.
The free membrane case was originally considered by Saffman and 
Delbr\"uck~\cite{saffman-75,saffman-76}, and the relevant length scale 
is the Saffman-Delbr\"uck length $\eta/\eta_{\rm s}$ beyond which the 
membrane feels the presence of the outer solvent.
On the other hand, the confined membrane case corresponds to that of a 
supported membrane close to the substrate as considered by Evans and 
Sackmann~\cite{evans-88} and later generalized by us~\cite{SRK11}.
Here the corresponding hydrodynamic screening length is set 
by the geometric mean of the Saffman-Delbr\"uck length $\eta/\eta_{\rm s}$
and the distance $h$ between the membrane and the wall, i.e., 
$\sqrt{\eta h/\eta_{\rm s}}$~\cite{SA98}.
Notice that the presence of the second wall only doubles the 
screening effect.
In the following, we shall mainly consider the confined membrane case
which allows us to treat most of the calculations analytically.   
This is mainly because $\kappa[\omega]$ does not depend on $k$.
However, it should be noted that the limiting expression of 
eq.~(\ref{conmem}) gives a reasonable approximation to the full 
expression of eq.~(\ref{genoseen}) even for $h \sim \eta/\eta_{\rm s}$
at least for a large enough moving object~\cite{SA98,RKSI11}.

Concerning the viscoelasticity of the surrounding solvent, we assume 
that its complex modulus obeys the power-law behavior such that 
$G_{\rm s}[\omega]=G_0(i \omega)^{\alpha}$ with $\alpha < 1$, as generally 
argued by Granek~\cite{granek-11}.
This behavior is commonly observed for various polymeric solutions at 
high frequencies.   
Examples are $\alpha=1/2$ and $\alpha=2/3$ for Rouse and Zimm 
dynamics, respectively~\cite{doiedwards}, $\alpha = 3/4$ for semi-dilute 
solutions of semi-flexible polymers such as actin filaments~\cite{GSOMS}.
From the viewpoint of particle-tracking microrheology 
experiment~\cite{Mason00,SM10}, 
it is more convenient to work in the Laplace domain defined by 
$\tilde{f}(s)= \int_0^{\infty} {\rm d} t \, f(t) e^{-st}$. 
Since we have been working in the Fourier-Laplace domain, it is 
straightforward to convert to the Laplace domain by substituting 
$s=i \omega$.
From $\tilde{G}_{\rm s}(s)=G_0 s^\alpha$, the Laplace transform of the solvent 
viscosity behaves as   
$\tilde{\eta}_{\rm s}(s) = \tilde{G}_{\rm s}(s)/s =G_0 s^{\alpha-1}$.
Then eq.~(\ref{sqscreen}) simply becomes 
\begin{equation}
\tilde{\kappa}^2(s) =  
\frac{2G_0 s^{\alpha-1}}{\eta h} + \frac{\rho s}{\eta}
\approx \frac{2G_0 s^{\alpha-1}}{\eta h}, 
\label{kappagen}
\end{equation}
where we have dropped the inertial term at the end.
This approximation is justified when $G_0 s^{\alpha-2}/\rho h \gg 1$, 
which is always valid for $s \rightarrow 0$ corresponding to the 
long-time behavior.

\section{Single-particle tracking}

For the description of the Brownian motion of a circular disk confined 
in a membrane, we basically follow the formulation in ref.~\cite{seki-93}. 
Let $a$ and $m$ be the radius and the mass of the disk, respectively.
The effective generalized Langevin equation is written 
as~\cite{KuboBook} 
\begin{equation}
m^{\ast} \frac{\rm d}{{\rm d}t}U(t)= 
-\int_{-\infty}^{t} {\rm d}t' \, 
\lambda(t-t')U(t') + R(t),
\end{equation}
where $U(t)$ is the velocity of the disk, and $\lambda(t)$ is the 
time-dependent drag coefficient given below.
The random force $R(t)$ satisfies the fluctuation dissipation 
theorem (FDT) when averaged over the ensemble of molecular motions; i.e.,
$\langle R(t) \rangle = 0$ and 
$\langle R(t_0)R(t_0+t) \rangle = k_{\rm B} T \lambda(t)$,
where $k_{\rm B}$ is the Boltzmann constant and $T$ the temperature.
In ref.~\cite{seki-93}, it was shown that the renormalized mass is given by 
$m^{\ast} = m + \pi \rho a^2$ which takes into account the additional inertia 
due to the drag from the fluid membrane.
Furthermore, the Laplace transform of $\lambda(t)$ is calculated to be
\begin{equation}
\tilde{\lambda}(s)=\frac{2 \pi \tilde{\eta}_{\rm s}(s) a^2}{h}
+ 4\pi \eta \frac{\tilde{\kappa}(s)a K_1[\tilde{\kappa}(s)a]}
{K_0[\tilde{\kappa}(s)a]},
\label{generaldrag}
\end{equation}
where $K_0[x]$ and $K_1[x]$ are modified Bessel functions of the second kind, 
order zero and one, respectively.

Following the standard procedure to obtain the MSD~\cite{KuboBook}, 
one can relate its Laplace transform and $\tilde{\lambda}(s)$ through 
\begin{equation}
\langle \Delta \tilde{\mathbf{r}}^2(s) \rangle
= \frac{4 k_{\rm B}T}{s^2 \tilde{\lambda}(s)},
\label{msdlaplace}
\end{equation}
where the number of degrees of freedom tracked in the MSD is chosen to 
be two.
For general $t$, application of the inverse Laplace transform provides
us with the time-dependent MSD;
\begin{equation}
\langle \Delta \mathbf{r}^2(t) \rangle = 
\frac{1}{2 \pi i}
\int_{c-i\infty}^{c+i\infty} {\rm d}s \, 
\frac{4 k_{\rm B}T}{s^2 \tilde{\lambda}(s)} e^{st}.
\label{msdtime}
\end{equation}

In order to demonstrate how to use the above relations, we consider here 
the limit of a large disk size $a \rightarrow \infty$ 
($\tilde{\kappa}(s)a \gg 1$)  
so that the drag coefficient in eq.~(\ref{generaldrag}) is dominated by 
the first term, i.e., 
$\tilde{\lambda}(s) \approx 2 \pi \tilde{\eta}_{\rm s}(s) a^2/h$.
Then the Laplace transformed MSD in eq.~(\ref{msdlaplace}) becomes
\begin{equation}
\langle \Delta \tilde{\mathbf{r}}^2(s) \rangle = 
\frac{2 k_{\rm B}T h}{\pi a^2 s^2 \tilde{\eta}_{\rm s}(s)} =
\frac{2 k_{\rm B}T h}{\pi a^2 s \tilde{G}_{\rm s}(s)}.
\label{msdmodulus}
\end{equation}
This is the equation which relates the observed 2D MSD to the modulus 
of the surrounding bulk solvent (rather than the membrane). 
In other words, we can extract the solvent 3D information by using    
the 2D information due to the motion of a disk in the membrane.
Once $\tilde{G}_{\rm s}(s)$ is obtained from the experiment, the 
frequency dependence of the storage and the loss moduli can be deduced
by identifying 
$G_{\rm s}[\omega]=G'_{\rm s}[\omega]+i G''_{\rm s}[\omega]
=\tilde{G}_{\rm s}(s=i \omega)$.
Notice that these two representations are equivalent because 
$G'_{\rm s}[\omega]$ and $G''_{\rm s}[\omega]$ are related
by the Kramers-Kronig relation~\cite{Mason00,SM10,KuboBook}.

On the other hand, suppose the modulus of the bulk solvent is known 
a priori to behave as $\tilde{G}_{\rm s}(s)=G_0 s^\alpha$ from  
independent experiments, the MSD can be simply obtained from 
eqs.~(\ref{msdtime}) and (\ref{msdmodulus}) as  
\begin{equation}
\langle \Delta \mathbf{r}^2(t) \rangle = 
\frac{2 k_{\rm B}T h}{\pi a^2 G_0 \Gamma[1+\alpha]}t^{\alpha}, 
\label{msdsubdif}
\end{equation}
where $\Gamma[x]$ is the gamma function.
This calculation shows that the viscoelasticity of the solvent results
in a subdiffusive time dependence of the MSD.
Since $\alpha <1$, the viscoelasticity slows down the normal diffusion 
process.
This is the main result of this Letter.
Compared with the 3D case~\cite{MW95,MGZWK}, the above expression is 
unique because it is proportional to $h/a^2$. 
This $1/a^2$-dependence arises from the mass conservation in 2D
rather than the momentum conservation~\cite{sanoop-poly-11}.

In the limit of a small disk size $a \rightarrow 0$ ($\tilde{\kappa}(s)a \ll 1$), 
the situation is more complicated.  
In this case, the drag coefficient asymptotically behaves as 
\begin{equation}
\tilde{\lambda}(s) \approx 
4 \pi \eta \left[ \ln \left( \frac{2}{\tilde{\kappa}(s)a} \right) 
- \gamma \right]^{-1},
\label{siglesmalldrag}
\end{equation}
where $\gamma=0.5772 \cdots$ is Euler's constant.
By using eq.~(\ref{kappagen}) for $\tilde{\kappa}(s)$, 
a similar calculation yields 
\begin{equation}
\langle \Delta \mathbf{r}^2(t) \rangle \approx 
\frac{k_{\rm B}T}{2\pi\eta} t 
\left[ \ln \left( \frac{2 \eta h t^{\alpha-1}}{G_0 a^2} \right)
+(\alpha -3) \gamma - \alpha +1 \right].
\label{singlemsdsmalldistance1}
\end{equation}
Since $\alpha < 1$, this MSD grows like $t \ln (1/t)$.
Such a logarithmic correction leads to a time-dependent diffusivity.
It should be noted, however, that this asymptotic expression is valid 
only when $G_0 a^2 t^{1-\alpha}/\eta h \ll 1$.
We also mention that all the above expressions recover our previous 
results when the solvent is purely viscous, i.e., 
$\alpha \rightarrow 1$~\cite{SRK11}.

\begin{figure}[t]
\begin{center}
\includegraphics[scale=0.4]{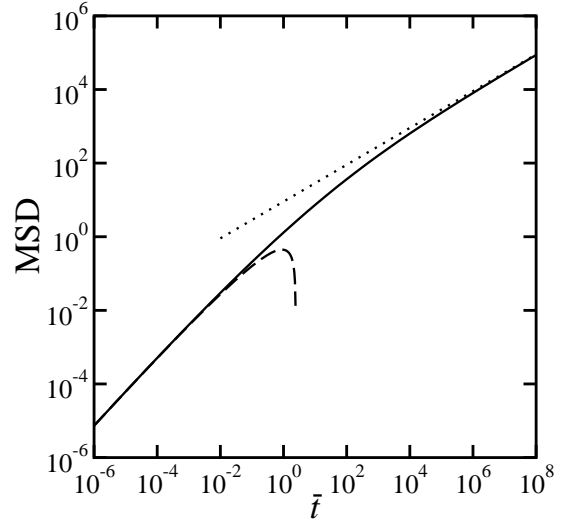}
\caption{Scaled MSD as a function of scaled time $\bar{t}$ 
when $\alpha=1/2$.  
These dimensionless quantities are defined in the text.
The dotted and dashed lines correspond to the asymptotic expressions
given by eqs.~(\ref{msdsubdif}) and (\ref{singlemsdsmalldistance1}), 
respectively.}
\label{fig2}
\end{center}
\end{figure}

In fig.~\ref{fig2}, we plot the result of numerical inverse Laplace 
transform of eq.~(\ref{msdtime}) with the full drag coefficient 
eq.~(\ref{generaldrag}) when $\alpha=1/2$, and compared it with the asymptotic 
expressions; eqs.~(\ref{msdsubdif}) and (\ref{singlemsdsmalldistance1}).
Here the dimensionless MSD and time are given by
$(2\pi\eta/k_{\rm B}T) (\eta h/2 G_0 a^2)^{1/(\alpha-1)}
\langle \Delta \mathbf{r}^2(t) \rangle$ and 
$\bar{t}= (\eta h/2 G_0 a^2)^{1/(\alpha-1)} t$, respectively.
As mentioned above, eq.~(\ref{singlemsdsmalldistance1}) is valid only for short 
time ($\bar{t} \ll 1$), while eq.~(\ref{msdsubdif}) describes the asymptotic 
long-time behavior ($\bar{t} \gg 16^{1/(1-\alpha)}$). 
The intermediate time region is described by an apparent power-law
$\bar{t}^{(1+\alpha)/2}$, which will be explained elsewhere.

\section{Two-particle tracking}

So far, we have discussed the motion of a single disk of finite 
radius $a$.
As discussed in ref.~\cite{SM10}, there are several advantages to perform 
multi-particle microrheology.
For example, long-time convective drift can be automatically subtracted
in this method so that measurements of probe self-diffusivities become 
possible over longer times. 
Multi-particle techniques can be also used to investigate heterogeneous 
materials. 
Here we discuss the effects of the solvent viscosity on the cross-correlation
of two distinct probe positions, namely, two-point 
microrheology~\cite{CVWGKYW}.
We show below that the distance between the two points essentially 
corresponds to the size of a disk in single-particle microrheology.

Consider a pair of point particles embedded in the membrane undergoing
Brownian motion separated by a 2D vector $\mathbf{r}$. 
The quantity of interest is the cross-correlation of the particle  
displacements 
$\langle \Delta r^1_\alpha(0) \Delta r^2_\beta(t)\rangle_r$,
where $\Delta r^i_\alpha$ is the displacement of the particle 
$i$ ($=1, 2$) along the axis $\alpha$ ($=x, y$).
We also define the $x$-axis to be along the line connecting the two 
particles, i.e., $\mathbf{r}=r \hat{\mathbf{e}}_x$.
According to the FDT, this correlation function is related to the coupling 
mobility $M_{\alpha \beta}(t)$ in the Laplace domain as~\cite{SM10}
\begin{equation}
\langle \Delta \tilde{r}^1_\alpha \Delta \tilde{r}^2_\beta(s) \rangle_r =
\frac{2 k_{\rm B}T}{s^2} \tilde{M}_{\alpha \beta}(r,s),
\label{2pMSD}
\end{equation}
for sufficiently large $r$.
The above equation is the analog of eq.~(\ref{msdlaplace}) for the 
two-particle tracking.
Since $M_{xy}=0$ by symmetry, we consider the longitudinal coupling 
mobility  $M_{\rm L}=M_{xx}$ and the transverse one $M_{\rm T}=M_{yy}$.
Note that the coupling mobility is not the inverse of the coupling 
resistance in the multi-particle case~\cite{SM10}.
By utilizing the results in refs.~\cite{RKG10,OD10,sanoop-poly-11},
one can obtain these mobilities analytically.

First the Laplace transform of the longitudinal coupling mobility turns 
out to be 
\begin{equation}
\tilde{M}_{\rm L}(r,s) = 
\frac{1}{2\pi\eta} \left[\frac{1}{(\tilde{\kappa}(s)r)^2} 
- \frac{K_1[\tilde{\kappa}(s) r]}{\tilde{\kappa}(s) r} \right].
\label{longmobility}
\end{equation}
In the limit of a large distance $r \rightarrow \infty$
($\tilde{\kappa}(s)r \gg 1$), 
the above expression can be approximated as
\begin{equation}
\tilde{M}_{\rm L}(r,s) \approx  
\frac{1}{2\pi\eta} \frac{1}{(\tilde{\kappa}(s)r)^2}. 
\label{longmobilitylarge}
\end{equation}
As in the calculation of MSD of a single disk, we use  
eq.~(\ref{kappagen}) for $\tilde{\kappa}(s)$ and perform the 
inverse Laplace transform of eq.~(\ref{2pMSD}). 
Then we obtain 
\begin{equation}
\langle \Delta r^1_x(0) \Delta r^2_x(t)\rangle_r \approx 
\dfrac{k_{\rm B}T h}{2\pi G_0 \Gamma[1+\alpha]} 
\dfrac{t^\alpha}{r^2}. 
\label{msdlargedistance1}
\end{equation}
The subdiffusive dependence on time and the $1/r^2$-dependence on 
distance is analogous to eq.~(\ref{msdsubdif}), implying that $r$ 
corresponds to $a$.

In the limit of a small distance $r \rightarrow 0$
($\tilde{\kappa}(s)r \ll 1$), 
on the other hand, eq.~(\ref{longmobility}) asymptotically behaves as  
\begin{equation}
\tilde{M}_{\rm L}(r,s) \approx 
\frac{1}{4\pi\eta}
\left[
\ln\left(\frac{2}{\tilde{\kappa}(s)r}\right)
-\gamma +\frac{1}{2}
\right].
\label{longmobilitysmall}
\end{equation}
Following the same process as above, the cross-correlation function 
asymptotically behaves as 
\begin{align}
\langle \Delta r^1_x(0) \Delta r^2_x(t)\rangle_r & \approx 
\frac{k_{\rm B}T}{4\pi\eta} t 
\biggl[ \ln \left( \frac{2 \eta h t^{\alpha-1}}{G_0 r^2} \right) 
\nonumber \\
& +(\alpha -3) \gamma - \alpha +2 \biggr],
\label{msdsmalldistance1}
\end{align}
which is valid for $G_0 r^2 t^{1-\alpha}/\eta h \ll 1$.
Notice that eq.~(\ref{msdsmalldistance1}) is also analogous to 
eq.~(\ref{singlemsdsmalldistance1}).

Next the transverse coupling mobility is given by
\begin{equation}
\tilde{M}_{\rm T}(r,s)= 
\frac{1}{2\pi\eta} \left[
K_0[\tilde{\kappa}(s) r]
+\frac{K_1[\tilde{\kappa}(s)r]}{\tilde{\kappa}(s) r} 
-\frac{1}{(\tilde{\kappa}(s) r)^2} 
\right].
\end{equation}
The large distance limit ($r \rightarrow \infty$) and the small distance 
limit ($r \rightarrow 0$) of this expression are    
\begin{equation}
\tilde{M}_{\rm T}(r,s) \approx  
-\frac{1}{2\pi\eta} \frac{1}{(\tilde{\kappa}(s)r)^2}, 
\end{equation}
and  
\begin{equation}
\tilde{M}_{\rm T}(r,s) \approx  
\frac{1}{4\pi\eta}
\left[
\ln\left(\frac{2}{\tilde{\kappa}(s)r}\right)
-\gamma  - \frac{1}{2}
\right],
\end{equation}
respectively.
Since these forms differ from eqs.~(\ref{longmobilitylarge}) and 
(\ref{longmobilitysmall}) only by a sign, we do not repeat here the  
same calculations.
However, as far as the time dependence of  
$\langle \Delta r^1_y(0) \Delta r^2_y(t)\rangle_r$  
is concerned, it is essentially given by 
eqs.~(\ref{msdlargedistance1}) and (\ref{msdsmalldistance1}).

\section{Discussion}

In this Letter, we have discussed the dynamics in a purely viscous 
lipid membrane surrounded by viscoelastic solvents such as polymeric
solutions. 
Using the generalized frequency-dependent mobility tensor for the 
confined membrane case, we calculated the MSD of a disk embedded in 
the membrane and obtained some asymptotic expressions.
The obtained MSD exhibits an anomalous diffusion reflecting the 
viscoelastic property of the bulk solvent. 
For single-particle microrheology experiments, we presented an 
useful relation which connects the MSD and the solvent modulus in 
the Laplace domain when the size of the disk is large enough.
We also obtained the cross-correlation of the particle displacements 
which can be used for two-particle tracking experiments.
Our theory can be applied not only for lipid membranes but also for
Langmuir monolayers.

It is worthwhile to point out the implicit assumptions which are used 
in the present theory~\cite{SM10}.
First we have assumed that the system obeys the FDT which allows us 
to relate the thermal fluctuations of the probe disk directly to the 
time-dependent drag coefficient (see eq.~(\ref{msdlaplace})) or the 
coupling mobility (see eq.~(\ref{2pMSD})).  
However, in non-equilibrium situations with membrane transport proteins,
for instance, the FDT can be violated and the deviation from the 
present theory may arise.
Recently, it was reported that an in vitro model system consisting 
of a cross-linked actin network with embedded force-generating 
myosin motors strongly violates FDT~\cite{Mizuno07}.
The second crucial assumption is that the generalized drag coefficient
or the coupling mobility are given precisely by their Newtonian 
analogs, but with the Newtonian viscosity $\eta_{\rm s}$ replaced
by the frequency-dependent complex viscosity $\eta_{\rm s}[\omega]$  
at all frequencies (see eq.~(\ref{eqn2:tau})).
This is not an obvious assumption, but it has proved to be quite 
successful for any probe motion in a simple linear viscoelastic
material under various conditions at least in 3D~\cite{SM10}.

Some caution is required when applying our theory to 
experiments if the viscoelastic solvent is, for instance, a 
semi-dilute polymer solution.    
For single-particle tracking, our continuum approach is valid 
for inclusion sizes much larger than the mesh size of the network.  
This can be relevant such as for a micron-size membrane domain and 
a network correlation length of tens of nanometers.
However our theory may not be applicable for small membrane proteins
which feel the solvent as purely viscous.
For two-particle tracking, the distance between the two point 
particles (which can be small membrane proteins) should be larger 
than the network mesh size. 
In order to obtain the full time behavior of the particle motion,
one should use a viscoelastic modulus that is dependent both 
on wavevector and frequency.
On the other hand, any simple fluid can be effectively viewed 
as viscoelastic at very high molecular frequencies~\cite{BZ}.
Our model can be also used to study such a high frequency dynamics 
by using a Maxwell model for the solvent.

In the present work, we have mainly discussed the confined membrane 
case in order to obtain analytical expressions.
Unfortunately, a single analytic expression of the drag coefficient for 
the whole range of the disk size is not known for the free membrane case. 
However, in the limit of a large disk size $a \rightarrow \infty$, 
the asymptotic expression was obtained by Hughes \textit{et al.} as 
$\lambda_{\rm free} \approx 16 \eta_{\rm s} a$ which depends only on the
solvent viscosity and is proportional to the size~\cite{hughes-81}. 
Assuming the replacement of the solvent viscosity with the time-dependent
one as discussed above, the MSD of the disk becomes
\begin{equation}
\langle \Delta \tilde{\mathbf{r}}^2(s) \rangle_{\rm free} = 
\frac{k_{\rm B}T}{4 a s^2 \tilde{\eta}_{\rm s}(s)} =
\frac{k_{\rm B}T}{4 a s \tilde{G}_{\rm s}(s)}.
\label{hughsmodulus}
\end{equation}
When the modulus of the bulk solvent behaves as 
$\tilde{G}_{\rm s}(s)=G_0 s^\alpha$ as before, the MSD in the 
free membrane case is given by 
\begin{equation}
\langle \Delta \mathbf{r}^2(t) \rangle_{\rm free} = 
\frac{k_{\rm B}T}{4 a G_0 \Gamma[1+\alpha]}t^{\alpha}, 
\label{msdsubdiffree}
\end{equation}
which is again a subdiffusive behavior.
In the opposite limit of $a \rightarrow 0$, the Laplace transformed drag 
coefficient can be written as~\cite{saffman-75,saffman-76} 
\begin{equation}
\tilde{\lambda}_{\rm free}(s) \approx 
4 \pi \eta \left[ \ln \left( \frac{\eta}{\tilde{\eta}_{\rm s}(s)a} \right) 
- \gamma \right]^{-1}.
\label{siglesmalldragfree}
\end{equation}
It then follows that  
\begin{equation}
\langle \Delta \mathbf{r}^2(t) \rangle_{\rm free} = 
\frac{k_{\rm B}T}{\pi\eta} t 
\biggl[ \ln \left( \frac{\eta t^{\alpha-1}}{G_0 a} \right)
+(\alpha -2) \gamma - \alpha +1 \biggr],
\label{singlemsdsmalldistance1free}
\end{equation}
for $G_0 a t^{1-\alpha}/\eta \ll 1$.
Although the above argument is not rigorous, it essentially captures the
anomalous diffusion behavior in the free membrane case.
For the two-particle tracking case, however, the corresponding cross-correlation
function can be obtained without any ambiguity since both the longitudinal 
and transverse coupling mobilities were analytically obtained for the free 
membrane case~\cite{OD10}.

Currently we are extending our model to the case when the two solvents 
have asymmetric viscoelastic properties or when the disk is actively
driven.
Additional effects such as the finite curvature of vesicles or the 
out-of-plane deformation of the membrane should also be taken into account.

\acknowledgments

We thank C.-Y.\ D.\ Lu for useful discussions. 
This work was supported by Grant-in-Aid for Scientific Research
(grant No.\ 21540420) from the MEXT of Japan.
SK also acknowledges the supported by the JSPS Core-to-Core Program
``International research network for non-equilibrium dynamics 
of soft matter''.


\end{document}